\documentclass[]{pasj02} 

\usepackage[switch,mathlines]{lineno} 

\usepackage{url}
\usepackage{lscape}
\usepackage{siunitx}
\usepackage{tabularx}
\usepackage{multirow}
\usepackage{booktabs}

\jyear{2024}
\Received{}
\Accepted{}

\begin{document} 

\title{Arcsecond-Scale X-ray Imaging and Spectroscopy of SS 433 with Chandra HETG}
\author{
Yusuke \textsc{Sakai},\altaffilmark{1}\altemailmark\orcid{0000-0002-5809-3516}
\email{sakai.yusuke.d@rikkyo.ac.jp} 
Shinya \textsc{Yamada},\altaffilmark{1}\orcid{0000-0003-4808-893X}
Haruka \textsc{Sakemi},\altaffilmark{2}\orcid{0000-0002-4037-1346}
Mami \textsc{Machida},\altaffilmark{3,4}\orcid{0000-0001-6353-7639}
Taichi \textsc{Igarashi},\altaffilmark{3,1}\orcid{0000-0003-4369-7314}
Ryota \textsc{Hayakawa},\altaffilmark{1,5}\orcid{0000-0002-3752-0048}
Miho \textsc{Tan},\altaffilmark{4,3}\orcid{0009-0000-2006-2688}
and
Taisei \textsc{Furuyama},\altaffilmark{1}
}


\altaffiltext{1}{Department of Physics, Rikkyo University, 3-34-1 Nishi Ikebukuro, Toshima-ku, Tokyo 171-8501, Japan}
\altaffiltext{2}{Graduate School of Science and Technology for Innovation, Yamaguchi University, 1677-1 Yoshida, Yamaguchi 753-0841, Japan}
\altaffiltext{3}{Division of Science, National Astronomical Observatory of Japan, 2-21-1 Osawa, Mitaka, Tokyo 181-8588, Japan}
\altaffiltext{4}{Astronomical Science Program, The Graduate University for Advanced Studies, SOKENDAI, 2-21-1, Osawa, Mitaka, 181-8588, Japan}
\altaffiltext{5}{International Center for Quantum-field Measurement Systems for Studies of the Universe and Particles (QUP), KEK, 1-1 Oho, Tsukuba, Ibaraki 305-0801, Japan}



\maketitle


\begin{abstract}
We present a spatial and spectral analysis of arcsecond-scale X-ray emission in SS 433 using zeroth-order data from Chandra High-Energy Transmission Grating (HETG) observations. The analysis is based on 24 observations acquired between 1999 and 2024, comprising a total exposure of $\sim$~850~ks and covering a wide range of orbital and precessional phases. Among these, the $\sim$~140~ks observation from 2014 was analyzed in detail for this study. This data provides the best statistics and was taken when the jets were nearly perpendicular to the line of sight and the accretion disk was eclipsed. By applying an energy-dependent subpixel event repositioning algorithm and the Richardson--Lucy deconvolution, we enhanced the spatial resolution and revealed eastern and western knot-like structures at a distance of $\sim$1$\farcs$7 ($\sim 10^{17}$~cm) from the core. These features are consistent with the kinematic precession model, and the positions of the knots suggest that they were ejected approximately 200 days prior to the observation. A comparison with VLA radio data obtained at a similar precessional phase shows that the X-ray emission extends east--west on a scale comparable to that of the radio emission. While the core is bright in both X-rays and radio, the brightness contrast between the knots and the core is smaller in X-rays than in radio. Spatially resolved spectroscopy indicates that prominent Fe lines in the core X-ray spectrum are well explained by thermal plasma emission. In contrast, Fe lines are not evident in the outer regions after accounting for potential core contamination, suggesting a dominant contribution from non-thermal processes. These findings imply that the arcsecond-scale X-ray structures may vary observationally with viewing conditions or precessional phase, but likely reflect a relatively stable jet-driving mechanism operating within the SS~433 system. 
\end{abstract}


\section{Introduction}

SS~433 remains one of the most studied jet sources in our Galaxy. It is an X-ray binary system consisting of an A-type supergiant star and a compact object (either a black hole or a neutron star) with an orbital period of 13.1~days \citep{Crampton_1980,Hillwig_2004}. The system, located at a distance of $\sim$~5.5~kpc, is known to eject relativistic jets moving at $0.26c$ ($c$ is the speed of light) \citep{blundell2004symmetry}. On larger spatial scales, the jets are associated with the W50 nebula, an elongated radio and X-ray structure that spans several degrees \citep{Dubner_1998, Watson_1983, Brinkmann_1996}.These jets exhibit precession with a period of 164~days, an opening half-angle of $20^\circ$, and an inclination angle of $79^\circ$, as inferred from Doppler-shifted H$\alpha$ line observations \citep{Abell_1979, Margon_1984}. This kinematic model is further supported by radio observations that reveal a helical jet structure extending to scales of $\sim10^{17}$~cm \citep{hjellming1981analysis, Spencer_1984}.
The jets carry a kinetic energy of $\sim 10^{39} \mathrm{erg\, s^{-1}}$, yet only $\sim 0.01\%$ of this energy ($\sim 10^{35} \mathrm{erg\, s^{-1}}$) is emitted as X-rays (see the review by \cite{fabrika_2006}). The remaining ``missing'' energy is thought to be transferred to the surrounding medium, potentially shaping the structure of the W50 radio nebula (e.g., \cite{Farnes_2017, Sakemi_2018, Ohmura_2021, Hayakawa_2022, Sakemi_2023}), although the exact mechanisms behind this energy transfer remain unclear.

Early X-ray CCD spectroscopy with ASCA revealed highly ionised, Doppler-shifted emission lines from Fe, Ni, and lighter elements \citep{Kotani_1994}. Subsequent observations with the Chandra High-Energy Transmission Grating (HETG; \cite{Canizares_2005}) observations helped establish SS~433 as the prototypical baryonic jet system \citep{Marshall_2002}. The X-ray--emitting region of the jet is located at $\sim10^{12}$~cm \citep{Kawai_1989, Brinkmann_1991, Marshall_2013}. The emission is dominated by an optically thin thermal plasma and is reasonably well reproduced by multi-temperature jet models that include both adiabatic and radiative cooling \citep{Marshall_2013, Khabibullin_2016, Medvedev_2019}. 

At larger scales ($\sim10^{17}$~cm), Chandra observations with the Advanced CCD Imaging Spectrometer (ACIS) and the zeroth-order imaging of the HETG have revealed arcsecond-separated structures located $\sim 1''$--$2''$ to the east and west of the core \citep{migliari2002iron, migliari2005rapid}. \citet{migliari2002iron} reported that these emission peaks are accompanied by Doppler-shifted Fe lines, which were interpreted as indicative of \textit{in situ} reheating in the jet flow. Based on spectral analysis of Doppler-shifted Fe lines, \citet{Khabibullin_2017} further suggested the presence of diffuse, non-spot-like emission extending over $\sim6 \times 10^{16}$~cm. Spectral analysis of this extended component showed a photon index of $\alpha \sim 0.7$ ($S_\nu \propto \nu^{-\alpha}$), consistent with radio observations \citep{Stirling_2004, Bell_2011}, implying a combination of thermal and non-thermal plasma components \citep{Miller_2008}.
Supporting the presence of non-thermal components, TeV $\gamma$-ray emission from the W50 lobes has been detected by HAWC, with subsequent confirmation by Fermi-LAT and H.E.S.S \citep{Abeysekara_2018,Li_2020,HESS_2024}.  
These $\gamma$-ray detections have motivated the development of various theoretical models \citep{sudoh_2020, kimura_2020} and X-ray studies \citep{Kayama_2022,Kayama_2025} aiming to investigate particle acceleration processes in the system.

Despite these advances, interpreting the arcsecond-scale X-ray structures and their emission properties remains nontrivial and needs careful consideration.
The angular scale of interest is so small that bright core emission (from $\sim 10^{12}$~cm) leaks into the outer jet regions at distances of $\sim 10^{17}$~cm, even within Chandra's superb angular resolution of $\sim 0\farcs5$.
To properly analyze the outer jet structures, it is essential to account for contamination from the bright core by considering the point-spread function (PSF).
However, ACIS data suffer from severe pile-up due to the high count rate from the core \citep{migliari2002iron, migliari2005rapid}, introducing additional uncertainty in estimating the core’s contribution to the outer jet emission.
Additionally, the X-ray flux from the core varies by up to an order of magnitude depending on the orbital and precessional phases of the system \citep{Marshall_2013, Medvedev_2019}.
These instrumental effects and intrinsic variability can complicate efforts to consistently isolate the outer jet emission.

In this study, we present a systematic analysis of all available HETG observations of SS~433, aiming to investigate the morphology and spectral properties of its X-ray jets.  
While grating insertion reduces the effective area by a factor of a few, the resulting decrease in photon flux serves to mitigate CCD pile-up from the bright core, making it a valuable trade-off in HETG observations.
Moreover, with over 20 HETG observations accumulated over the past two decades, SS~433 provides a uniquely rich dataset for studying the east--west jet emission and its dependence on orbital and precessional phases.
Spatially resolved zeroth-order spectra prove useful for exploring the emission from the outer jets, and the HETG's high energy resolution provides complementary constraints on the spectral modeling.

This paper is structured as follows.
Section~\ref{Observation and Data Analysis} describes the observations and data processing.
Section~\ref{Results} presents the imaging and spectral analysis of the arcsecond-scale jets.
In Section~\ref{Discussion}, we interpret the results in the context of the precession model and discuss consistency with VLA radio data and spectral properties.
Finally, Section~\ref{Conclusion} summarizes our key findings.

\begin{table}[htbp]
\caption{Ephemeris and Precession Model Parameters for SS~433}
\label{Ephemeris for SS 433}
\small
\begin{tabular}{@{\extracolsep{\fill}}l|ccc}
\toprule
Category & Parameter & Value & Ref. \\ \midrule[\heavyrulewidth]
\multirow{2}{*}{Period} 
    & $P_\mathrm{orb}$ & $13.08223$ d & 1 \\
    & $P_\mathrm{prec}$ & $162.15$ d & 2 \\ \midrule
\multirow{2}{*}{Reference epoch (HJD)} 
    & $t_{\mathrm{ref,orb}}$ & $2450023.746$ & 1 \\
    & $t_{\mathrm{ref,prec}}$ & $2451458.12$ & 2 \\ \midrule
Jet speed & $v$ & $0.2602c$ & 2 \\
Half-opening angle & $\theta$ & $19.85^\circ$ & 2 \\
Inclination angle & $i$ & $78.83^\circ$ & 2 \\
Position angle & $\chi$ & $98.2^\circ$ & 3 \\
Sense of rotation & $s_{\mathrm{rot}}$ & $-1$ & 4 \\
Distance & $d$ & $5.5$ kpc & 5 \\ \bottomrule
\end{tabular}

\vspace{0.3cm}
\small{
\textbf{References.}---(1) \citet{goranskij2011photometric}; (2) \citet{Gies_2002}; (3) \citet{Stirling_2002}; (4) \citet{hjellming1981analysis}; (5) \citet{blundell2004symmetry,Lockman_2007}.
}
\end{table}

\begin{figure}[ht!]
 \includegraphics[width=1\linewidth]{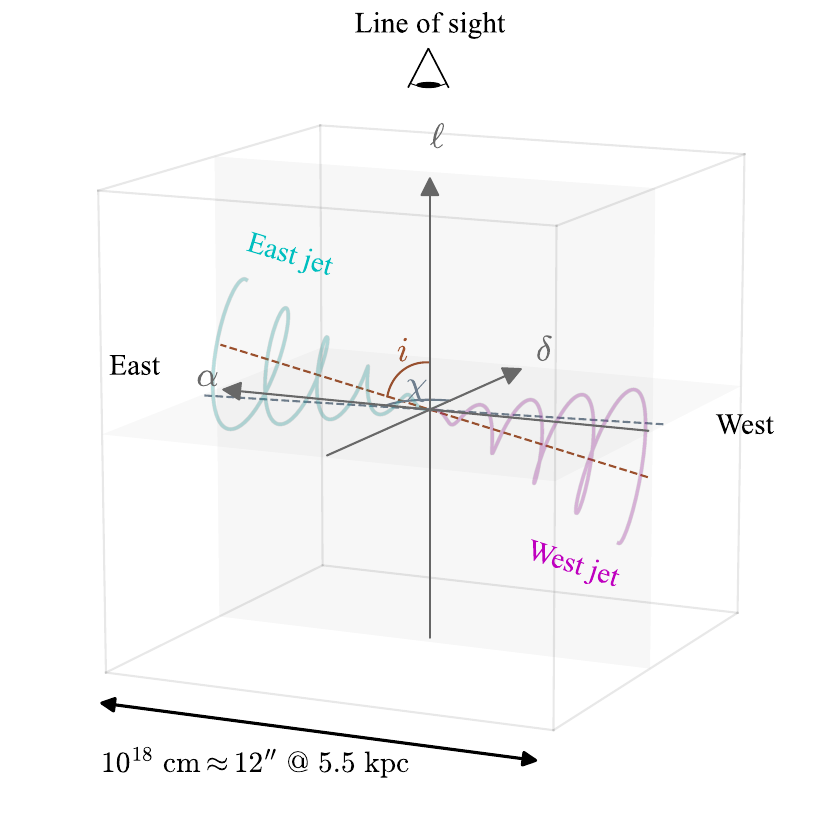}
\caption{Three-dimensional visualization of kinematic precession model of SS~433, constructed using parameters listed in Table~\ref{Ephemeris for SS 433}.  
Axes $\alpha$, $\delta$, and $\ell$ correspond to right ascension, declination, and line-of-sight direction, respectively.  
Cyan and magenta lines trace eastern and western jets over 4.5 precession cycles.  
Bounding box indicates physical scale of $10^{18}$~cm.
}
 \label{jet_model_3d}
\end{figure}

\section{Observation and Data Analysis}\label{Observation and Data Analysis}

\begin{table*}[ht!]
 \caption{Basic Information on the Chandra HETG Observations of SS 433 Used in this Paper}
   \begin{tabular*}{\textwidth}{@{\extracolsep{\fill}}rcrrrrrr}
   \toprule
      Obs. ID & Obs. Start & Exp. Time &  Frame Rate & Orbital Phase$^1$ & Precession Phase$^2$  & Energy Flux$^3$ & Direction$^4$  \\
       & (yyyy~mmm~dd) & (ks) & (s) & $\phi_\mathrm{orb}$ & $\phi_\mathrm{prec}$ & $F_\mathrm{HEG}$ & (degree)  \\\midrule[\heavyrulewidth]
106 & 1999 Sep 23 & 28.7 & 3.2 & 0.64 & 0.92 & 0.9 & 1 \\
1020 & 2000 Nov 28 & 22.7 & 2.5 & 0.67 & 0.58 & 0.3 & 326 \\
1019 & 2001 Mar 16 & 23.4 & 2.5 & 0.95 & 0.25 & 0.9 & 183 \\
1940 & 2001 May 08 & 19.6 & 3.2 & 0.97 & 0.58 & 0.2 & 158 \\
1941 & 2001 May 10 & 18.5 & 3.2 & 0.12 & 0.59 & 0.1 & 157 \\
1942 & 2001 May 12 & 19.7 & 3.2 & 0.27 & 0.60 & 0.6 & 156 \\
5512 & 2005 Aug 06 & 19.7 & 1.8 & 0.52 & 0.14 & 1.7 & 38 \\
5513 & 2005 Aug 12 & 48.1 & 1.8 & 0.97 & 0.18 & 1.0 & 31 \\
5514 & 2005 Aug 15 & 73.1 & 1.8 & 0.21 & 0.20 & 1.5 & 25 \\
6360 & 2005 Aug 17 & 57.4 & 1.8 & 0.40 & 0.21 & 1.5 & 25 \\
15781 & 2014 Aug 09 & 138.2 & 1.8 & 0.01 & 0.43 & 0.2 & 34 \\
20132 & 2018 Aug 10 & 18.6 & 1.7 & 0.77 & 0.45 & 1.0 & 33 \\
20131 & 2018 Aug 13 & 94.3 & 1.7 & 0.01 & 0.47 & 0.8 & 27 \\
24702 & 2021 May 19 & 29.4 & 1.8 & 0.22 & 0.70 & 0.8 & 152 \\
25047 & 2021 May 21 & 10.5 & 1.8 & 0.34 & 0.71 & 0.5 & 152 \\
25048 & 2021 May 22 & 30.4 & 1.8 & 0.41 & 0.71 & 0.2 & 152 \\
24703 & 2021 Jul 03 & 33.9 & 1.7 & 0.67 & 0.97 & 1.6 & 90 \\
25081 & 2021 Jul 04 & 32.9 & 1.7 & 0.73 & 0.98 & 1.6 & 90 \\
24704 & 2021 Aug 18 & 38.6 & 1.7 & 0.18 & 0.26 & 1.5 & 24 \\
25690 & 2021 Aug 20 & 34.2 & 1.7 & 0.28 & 0.27 & 1.4 & 24 \\
27356 & 2024 Mar 06 & 9.9 & 1.7 & 0.35 & 0.00 & 0.4 & 186 \\
29311 & 2024 Mar 10 & 9.9 & 1.8 & 0.64 & 0.02 & 0.2 & 187 \\
27357 & 2024 May 22 & 19.6 & 1.7 & 0.21 & 0.47 & 0.7 & 149 \\
27358 & 2024 Aug 06 & 19.6 & 1.7 & 0.04 & 0.94 & 0.7 & 35 \\
\bottomrule
   \end{tabular*}
  \label{used_data}

 \vspace{0.3cm}
\small{
\textbf{Notes.}---
$^1$ Ephemeris at observation start time follows \citet{goranskij2011photometric}.  
$^2$ Ephemeris at observation start time follows \citet{Gies_2002}.  
$^3$ Absorbed flux in 2--7.5~keV, based on XSPEC fit with \texttt{tbabs*powerlaw} ($N_\mathrm{H} = 1 \times 10^{22}\, \mathrm{cm^{-2}}$; \cite{Marshall_2002}).  
Energy flux is given in units of $10^{-10} \mathrm{erg\, cm^{-2}\, s^{-1}}$.  
$^4$ Charge transfer direction is measured counterclockwise, with $0^\circ$ at west, $90^\circ$ at north, $180^\circ$ at east, and $270^\circ$ at south.
}
\end{table*}

\subsection{Overview and Kinematic Precession Model} \label{Overview and Kinematic Precession Model}

In order to interpret the observed structure of SS~433, we begin by introducing the kinematic precession model commonly used to describe its jet motion (e.g., \cite{Stirling_2002}).

The jet velocity components are defined as follows:
\begin{equation}
\begin{aligned}
\beta_\alpha  &= \pm \beta \left[ \sin\chi \left( \sin i \cos\theta - \cos i \sin\theta \cos\psi \right) - s_\mathrm{rot} \sin\theta \cos\chi \sin\psi \right], \\
\beta_\delta  &= \pm \beta \left[ \cos\chi \left( \sin i \cos\theta - \cos i \sin\theta \cos\psi \right) + s_\mathrm{rot} \sin\theta \sin\chi \sin\psi \right], \\
\beta_\ell    &= \pm \beta \left( \cos i \cos\theta + \sin i \sin\theta \cos\psi \right).
\end{aligned}
\label{velocity_components}
\end{equation}
The parameters used in this model are summarized in Table~\ref{Ephemeris for SS 433}. The plus and minus signs correspond to the eastern and western jets, respectively. $\beta = v/c$ is the jet velocity, with components $\beta_\alpha$, $\beta_\delta$, and $\beta_\ell$ corresponding to the directions of right ascension, declination, and line of sight ($\beta_\ell > 0$: approaching the observer), respectively. The precession angle is defined as $\psi = 2\pi \phi_{\mathrm{prec}}$, where $\phi_{\mathrm{prec}} = (t - t_{\mathrm{ref,prec}})/P_{\mathrm{prec}}$, and $t$ is the observation time in Heliocentric Julian Date (HJD).

Figure~\ref{jet_model_3d} presents a three-dimensional view of the SS~433 jet system, constructed using the velocity components defined in Equation~\eqref{velocity_components}.
The eastern jet (cyan) and the western jet (magenta) trace out helical paths over a span of 4.5 precession cycles. 
The eastern jet is, on average, directed toward us, while the western jet recedes.

Due to their relativistic speed ($v \approx 0.26c$; see Table~\ref{Ephemeris for SS 433}), the jets exhibit relativistic Doppler shifts in their emission lines (e.g., \cite{Eikenberry_2001}). 
These shifts are described by the following expression:
\begin{equation}
    z = \gamma(1-\beta_\ell) - 1 = \gamma \left[1 \mp \beta (\cos i \cos \theta + \sin i \sin \theta \cos \psi) \right] - 1,
\label{precession_eq}
\end{equation}
where $\gamma = (1 - \beta^2)^{-1/2}$ is the Lorentz factor. The minus and plus signs correspond to the eastern jet ($z_{\mathrm{b}}$) and western jet ($z_{\mathrm{r}}$), respectively (see Figure~\ref{orb_vs_jet_phase}(b) for the relation between precession phase and Doppler shift).
This kinematic model is utilized for the analyses in Section~\ref{Comparison of X-ray Images with Jet Model}.

\subsection{Chandra X-ray Observatory}
\begin{figure}[ht!]
 \includegraphics[width=1\linewidth]{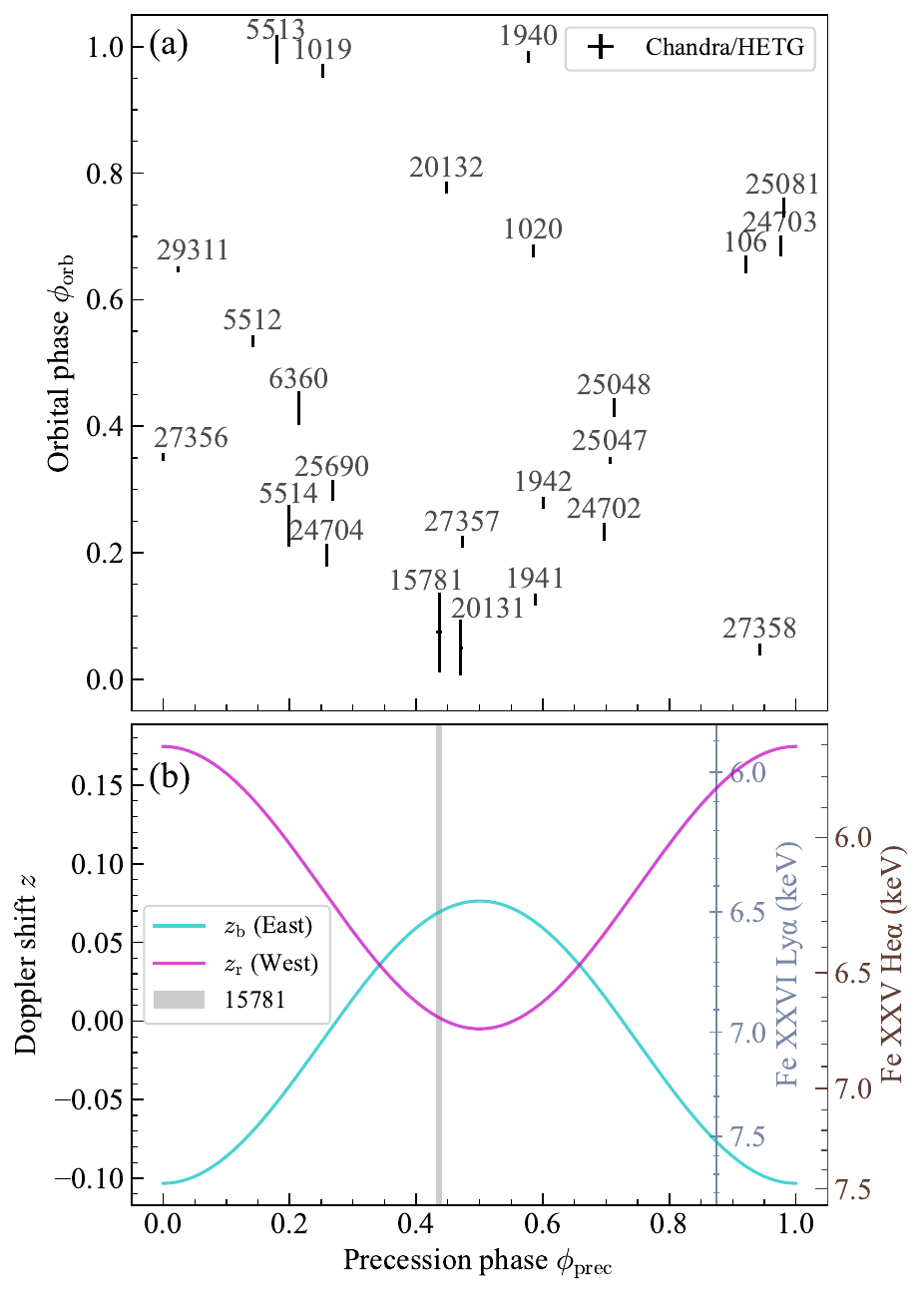}
\caption{(a) Relationship between precession phase ($\phi_{\mathrm{prec}}$) and orbital phase ($\phi_{\mathrm{orb}}$) observed with Chandra HETG.  
Error bars represent duration of each observation.  
(b) Predicted redshift variations based on relativistic precession motion of jets, calculated using Equation~\eqref{precession_eq}.  
Eastern jet ($z_{\mathrm{b}}$) and western jet ($z_{\mathrm{r}}$) are shown in cyan and magenta, respectively.  
Right-hand vertical axes indicate redshifted emission-line energies corresponding to Fe\,\textsc{xxvi} Ly$\alpha$ (rest-frame: 6.95~keV) and Fe\,\textsc{xxv} He$\alpha$ (rest-frame: 6.70~keV) transitions.  
Observation period of Obs.~ID~15781 is indicated by gray shading.
}
 \label{orb_vs_jet_phase}
\end{figure}

SS~433 has been observed 24 times with the HETG between late 1999 and 2024, with a total exposure time of $\sim$~850~ks. Table~\ref{used_data} summarizes all available observations, and Figure~\ref{orb_vs_jet_phase} shows their orbital and precession phase coverage. The observations cover a wide range of orbital and precession phase combinations.
The orbital phase is defined as $\phi_{\mathrm{orb}} = (t - t_{\mathrm{ref,orb}})/P_{\mathrm{orb}}$, where $\phi_{\mathrm{orb}} = 0$ corresponds to the eclipse (i.e., when the optical star passes in front of the accretion disk). The jet precession phase $\phi_{\mathrm{prec}}$, as defined in Section~\ref{Overview and Kinematic Precession Model}, takes $\phi_{\mathrm{prec}} = 0$ when the jet axis is most closely aligned with our line of sight (i.e., the most inclined phase).
We focus on a systematic analysis of these HETG datasets in the following sections.
The data were processed using CIAO \citep{fruscione2006ciao} version~4.15 with CALDB version~4.10.4.

\begin{figure*}[ht!]
 \includegraphics[width=1\linewidth]{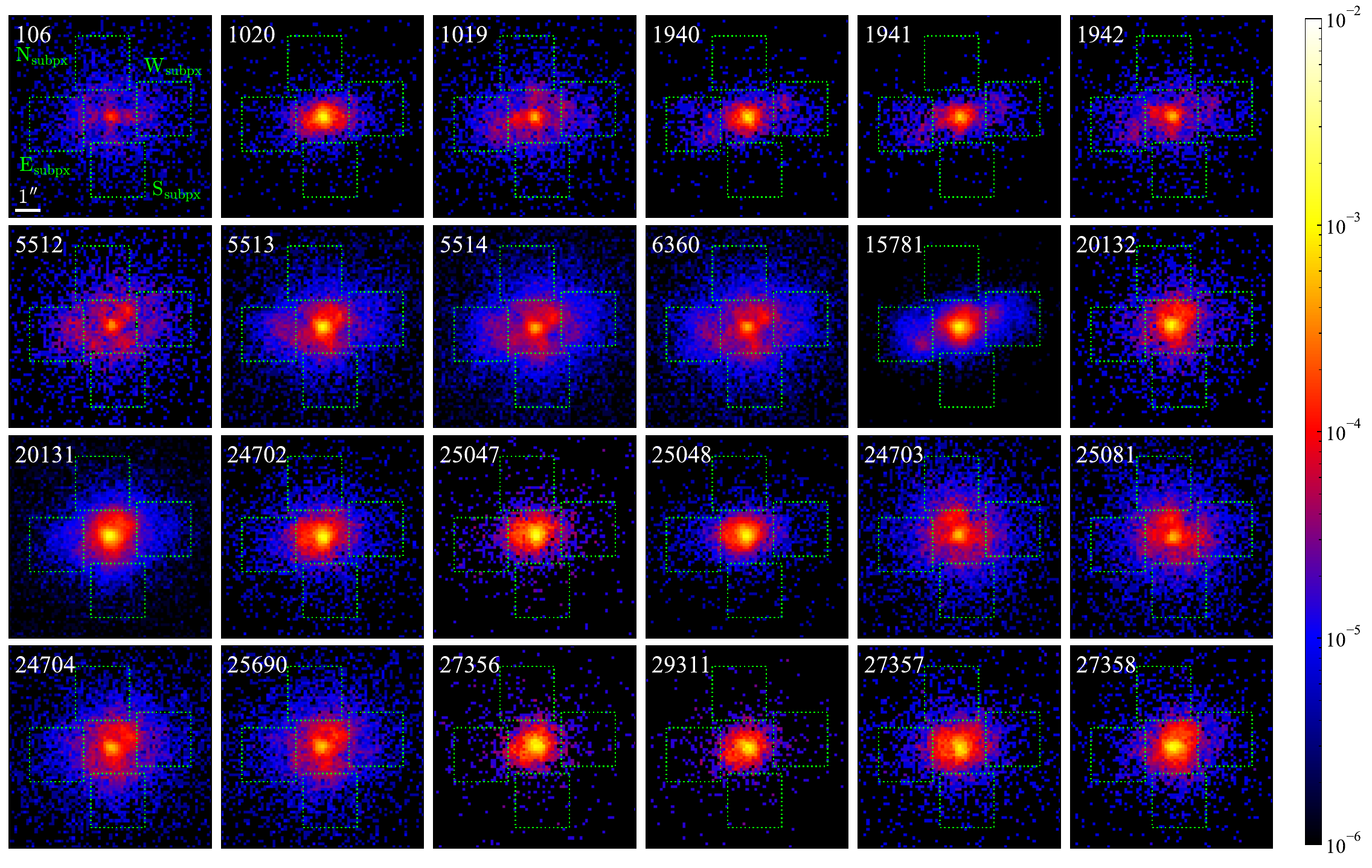}
\caption{Chandra HETG zeroth-order images in the 0.5--8~keV band, reconstructed at 1/4 pixel scale. Each panel corresponds to a different observation, with the Obs.~ID indicated at the top. The four rectangular regions---$ \mathrm{E_{subpx}}$, $ \mathrm{W_{subpx}}$, $ \mathrm{N_{subpx}}$, and $ \mathrm{S_{subpx}}$---are centered at positions offset by 19 subpixels from the brightest central pixel, with a rotation angle of 98.2$^\circ$ (see Table~\ref{Ephemeris for SS 433} for the position angle). Each region spans 19$\times$19 subpixels and is used for flux extraction in Figure~\ref{all_hetg_flux_ratio}. The flux units are given in \si{photons.cm^{-2}.\textit{detector}.\textit{pixel}.\textit{size}^{-2}.s^{-1}}. 
}
 \label{all_observation_hetg_subpixel}
\end{figure*}

\begin{figure}[ht!]
 \includegraphics[width=1\linewidth]{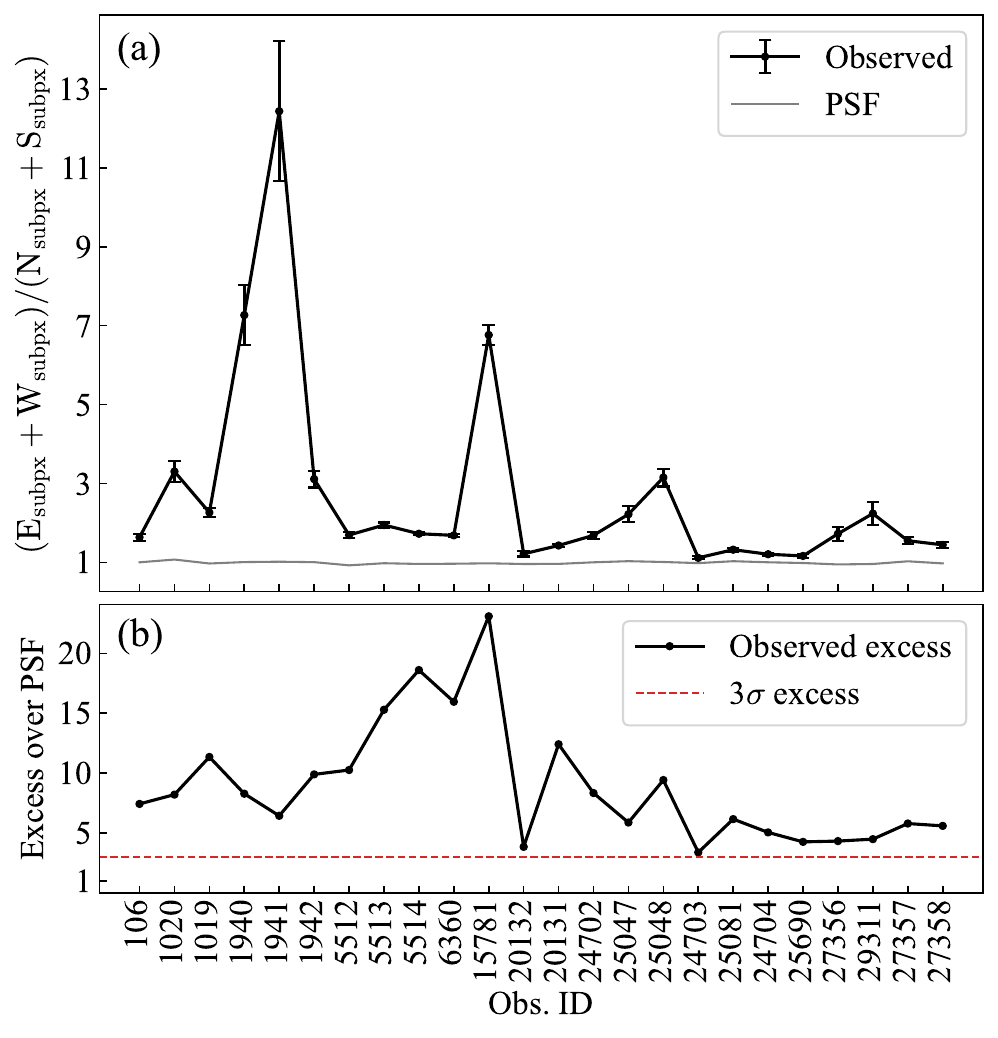}
 \caption{(a) Flux ratio between jet axis (E$_\mathrm{subpx}$ + W$_\mathrm{subpx}$) and orthogonal axis (N$_\mathrm{subpx}$ + S$_\mathrm{subpx}$), measured from 1/4-pixel resolution images in Figure~\ref{all_observation_hetg_subpixel}.  
Error bars represent statistical uncertainties.  
Gray plot shows expected ratio for a point source, calculated from 1/4-pixel PSF at 3.7~keV using identical region definitions.  
(b) Excess of observed flux ratio over corresponding PSF base value, normalized by statistical error.  
Horizontal dashed line indicates $3\sigma$ excess threshold.  
All observations lie above this threshold.
}
 \label{all_hetg_flux_ratio}
\end{figure}

\subsubsection{Imaging Analysis}

To explore the detailed spatial structure, we apply the energy-dependent subpixel event repositioning (EDSER; \cite{Li_2004}) algorithm. This method utilizes a charge diffusion model based on the detector's energy dependence to reconstruct the true photon impact positions with subpixel precision. As a result, it achieves spatial resolution finer than the detector's native pixel scale (e.g., \cite{Wang_2011}).

The data were reprocessed from the level 1 event file using \texttt{chandra\_repro} with the parameter \texttt{pix\_adj=EDSER} to apply the EDSER algorithm via CIAO. Exposure-corrected images were generated using \texttt{fluximage} over the 0.5--8~keV energy range with a pixel scale of \texttt{binsize=0.25} ($0\farcs123$). Throughout this paper, the \textit{detector pixel size} (\texttt{binsize=1}, $0\farcs492$) is used as the spatial unit for flux representation.

Although the EDSER algorithm improves spatial resolution by mitigating pixelation effects, it does not correct for the PSF. To further enhance image fidelity, we apply image deconvolution in conjunction with the EDSER algorithm. This approach has been effectively employed in previous studies, such as resolving fine jet structures in 3C~273~\citep{Marchenko_2017} and comparing with high-resolution radio observations in Pictor~A~\citep{Thimmappa_2020}.

Image deconvolution methods can be broadly categorized into those that incorporate prior information (e.g., \cite{Evans_1989, Esch_2004}) and those that rely solely on observational data (e.g., see the review by \cite{Starck_2002}). The Richardson--Lucy (RL; \cite{richardson_1972,lucy_1974}) algorithm belongs to the latter category, as it converges to the maximum likelihood solution under Poisson noise through iterative updates \citep{Shepp_1982}. The iterative rule is expressed as
\begin{equation}
W^{(r+1)} = W^{(r)}  \left[ \frac{N}{W^{(r)} * P} * P^\dagger \right],
\label{rl_eq}
\end{equation}
where $W^{(r)}$ is the reconstructed image at the $r$-th iteration, $N$ is the observed image, $P$ is the PSF, and $P^\dagger$ is the flipped PSF. The symbol $*$ denotes the convolution operator.

The RL method is well suited for X-ray data and has been widely used, owing to its compatibility with Poisson statistics (e.g., \cite{Grefenstette_2015, Sobolenko_2022, Sakai_2023, Morii_2024}). The number of iterations in the RL algorithm is not determined analytically and may vary depending on the data. In this study, we determine the optimal number based on the convergence behavior of the reconstructed image, utilizing the $\chi^2$ value and drawing on the approach used by \citet{Marchenko_2017}. Details are provided in Appendix~\ref{Optimal Number of Iterations for the RL Deconvolution}.

The PSF was simulated using MARX~\citep{davis2012raytracing} version 5.5.3 via CIAO's \texttt{simulate\_psf} tool, assuming a monoenergetic energy of 3.7~keV (the average photon energy in the dataset) and a pixel scale of \texttt{binsize=0.25}. For HETG observations, non-zeroth-order components were removed to isolate the zeroth-order image, and the PSF was normalized to unity. All operations in Equation~\eqref{rl_eq} were performed pixel-wise at a resolution of 1/4~pixel, using the EDSER-reconstructed image and the corresponding PSF.

For the imaging analysis, we initially examined all available HETG datasets to evaluate the data quality. Among them, Obs.~ID~15781 was selected as the primary target for detailed subpixel and deconvolution analyses, as it offers the longest exposure (138.2~ks) and adequate photon statistics. The orbital and precession phases during this observation were $\phi_{\mathrm{orb}}=0.01\text{--}0.14$ and $\phi_{\mathrm{prec}}=0.43\text{--}0.44$, respectively.

\begin{figure*}[ht!]
 \includegraphics[width=1\linewidth]{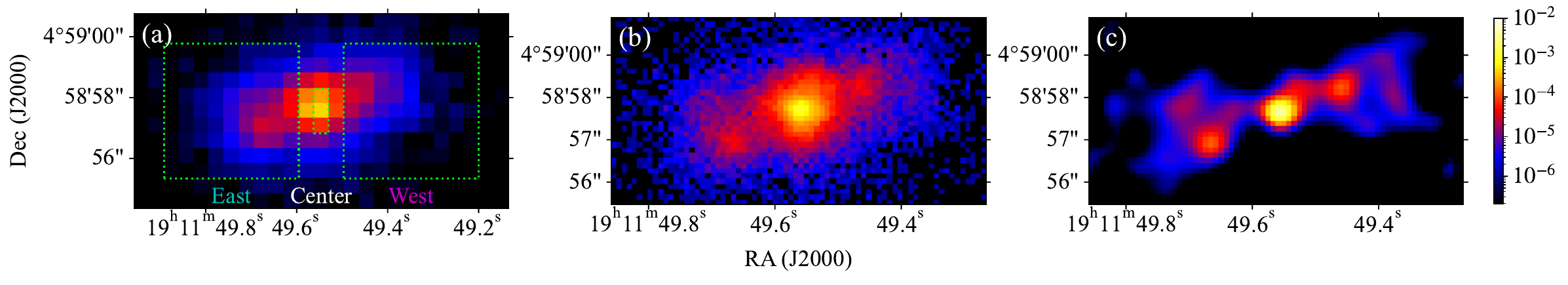}
\caption{Zeroth-order image of Obs.~ID~15781 (0.5--8~keV), observed in 2014 and displayed at the detector's native pixel scale of $0\farcs492$. Three rectangular regions denote the areas selected for spectral analysis. (b) Sub-pixel image (1/4 pixel scale) reconstructed using the EDSER algorithm. (c) RL-deconvolved image derived from (b). Flux units in each panel are \si{photons.cm^{-2}.\textit{detector}.\textit{pixel}.\textit{size}^{-2}.s^{-1}}. {Alt text: All panels are images with a color scale from 2e-7 to 1e2. In panel (a), the Center region is located at row 1, column 3. The East and West regions are centered at row 9, column 9, and offset by 6 native pixels to the east and west, respectively.}}
 \label{obs_subpixel_deconv}
\end{figure*}

\subsubsection{Spectral Analysis} \label{Spectral Analysis}
To investigate the spatially resolved spectral properties of SS~433, we analyzed zeroth-order spectra extracted from Obs.~ID~15781 in three regions of interest: Center, East, and West, defined based on features in the zeroth-order image with a pixel scale of \texttt{binsize=1} (see Figure~\ref{obs_subpixel_deconv}). We also analyzed the HETG spectra, which offer high spectral resolution and help constrain the spectral components in the zeroth-order spectra.

Zeroth-order spectra were extracted using the \texttt{specextract} script, with background spectra taken from regions located $10''$ north and south of the core. The HETG spectra were processed with \texttt{chandra\_repro} and combined with first-order ($\pm$1) HEG spectra using \texttt{combine\_spectra}. For the HETG data, background spectra were taken from the default background regions provided by \texttt{chandra\_repro}. The background contribution in the analyzed energy range is below $1\%$ for the Center, East, and West regions, and below $3\%$ for HEG. Spectral fitting and analysis were performed with XSPEC \citep{xspec_tool} version~12.14.

\subsection{Very Large Array}\label{Very Large Array}
SS 433 has been observed multiple times with the VLA. For comparison with the structure in the Chandra data of Obs.~ID~15781, we selected the VLA data from 2003 July 11, which consists of an 11-hour observation. Although no simultaneous observations were available and the orbital phase ($\phi_{\mathrm{orb}} = 0.63\text{--}0.66$) differs from that of the Chandra data, we utilized this dataset because it corresponds to a similar precession phase ($\phi_{\mathrm{prec}} = 0.470\text{--}0.473$). This VLA observation, conducted in the A configuration at C band (4--8 GHz), used 3C 286 as the primary flux calibrator and J1922+1530 as the gain calibrator. Data reduction was performed following standard procedures using CASA \citep{CASA_2022}, version~6.5.4.9. The process included bad data flagging, iterative self-calibration to enhance image quality, and deconvolution using the \texttt{tclean} task. To match the spatial resolution of the Chandra deconvolution image, the \texttt{imsmooth} task was applied. The resulting image is compared with the Chandra data in Section~\ref{Correlation of X-ray and Radio Images}.

\section{Results}\label{Results}

\subsection{Jet Axis Flux Asymmetry}

To investigate the extended emission from the jets, the flux along the jet axis and the orthogonal direction was analyzed.  
Figure~\ref{all_observation_hetg_subpixel} presents EDSER images (with \texttt{binsize=0.25}) from all HETG observations.
The regions labeled $ \mathrm{E_{subpx}}$, $ \mathrm{W_{subpx}}$, $ \mathrm{N_{subpx}}$, and $ \mathrm{S_{subpx}}$ were defined to evaluate asymmetry along the jet axis.  
Figure~\ref{all_hetg_flux_ratio}(a) shows the flux ratio between the jet axis ($\mathrm{E_{subpx}}$ + $\mathrm{W_{subpx}}$) and the orthogonal axis ($\mathrm{N_{subpx}}$ + $\mathrm{S_{subpx}}$), derived from the subpixel images shown in Figure~\ref{all_observation_hetg_subpixel}.

Due to small pointing offsets and variations in roll angle among the observations, the expected PSF ratio was computed at 3.7~keV using the same spatial resolution and region definitions.
In all cases, the observed flux ratios exceeded the PSF expectation by more than $3\sigma$, indicating that extended jet-like emission is consistently detected across all observations.
In addition, some observations (e.g., Obs.~ID~15781) were taken during phases when the emission is more prominent. The variations in visibility related to orbital and precession phases are discussed in Section~\ref{Phase-Dependent Jet Visibility}.

\subsection{Detailed Spatial Structure of the Jets}

For detailed image analysis, the RL deconvolution was applied to the EDSER-reconstructed data.
Figure~\ref{obs_subpixel_deconv}(a) shows the image at the detector's native pixel scale (\texttt{binsize=1}) from Obs.~ID~15781.  
Figures~\ref{obs_subpixel_deconv}(b) and (c) present the EDSER-reconstructed image at \texttt{binsize=0.25} and the RL-deconvolved image after 30 iterations, respectively.
The method for determining the optimal number of RL iterations is described in Appendix~\ref{Optimal Number of Iterations for the RL Deconvolution}.
In the deconvolved image (Figure~\ref{obs_subpixel_deconv}(c)), two bright structures appear at a distance of $\sim 1\farcs7$ from the core.
Moreover, the extended emission is not point-like but exhibits a structured morphology aligned with the jet axis.

\subsection{Comparison of X-ray Images with Jet Model} \label{Comparison of X-ray Images with Jet Model}

The deconvolved image (Figure~\ref{obs_subpixel_deconv}(c)) is compared with the kinematic model described by Equation~\eqref{velocity_components}. While the model provides a three-dimensional representation of the jet structure and velocity field, it does not account for the light-travel-time effect, which are essential when comparing with the observed X-ray images. Due to the finite speed of light and the spatial extent of the jets, the apparent positions of jet components are affected by the time it takes for photons to reach the observer. We therefore incorporate a light-travel-time effect into the model, as detailed below.

To define the birth epoch of each jet component, we use the reference time $t_0$ and the ejection time $t$, both expressed in HJD. $t_0$ corresponds to the midpoint of Obs.~ID~15781, and $t$ denotes the ejection time of the jet material. The birth epoch, defined as $t_{\mathrm{age}} = t_0 - t$ ($\geq 0$), represents the time elapsed since ejection and serves as a unique identifier for each jet component.

Due to the light-travel-time effect, the observed emission age $\tau$ differs from $t_{\mathrm{age}}$. Following \citet{roberts2010structure}, the relationship is given by
\begin{equation}
\tau = \frac{t_{\mathrm{age}}}{1 - \beta_\ell}.
\label{tau_vs_tage_eq}
\end{equation}
A visual representation of this relation is provided in Appendix~\ref{Light-Travel-Time Mapping of Jet Emission in SS 433}. As a result, the apparent position of the jet on the celestial sphere is calculated using
\begin{equation}
(\mathrm{RA},\, \mathrm{Dec}) = (\beta_\alpha c \tau,\, \beta_\delta c \tau).
\label{ra_dec_position}
\end{equation}

\begin{figure*}[ht!]
 \includegraphics[width=1\linewidth]{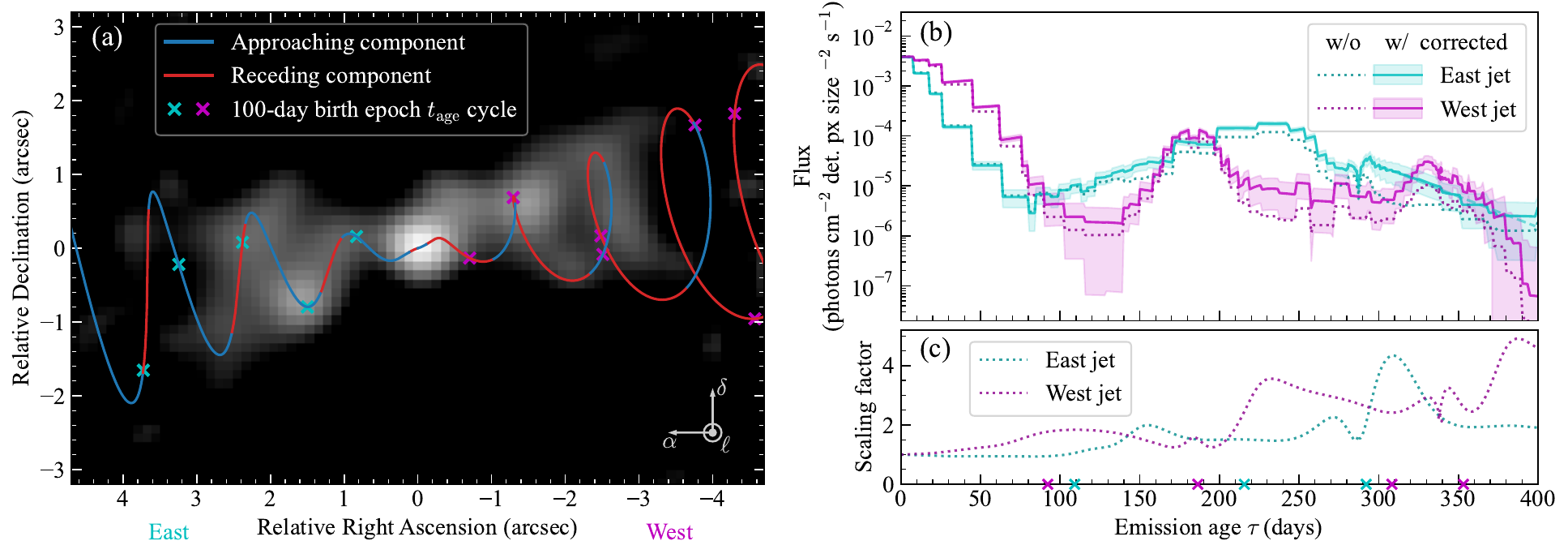}
 \caption{(a) RL-deconvolved image from Figure~\ref{obs_subpixel_deconv}(c), overlaid with the kinematic model. 
Approaching and receding jet components are shown in blue and red, respectively. 
Cyan and magenta crosses indicate birth epoch points $t_{\mathrm{age}}$ at 100-day intervals, corresponding to positions in panel (c). 
Axes at the bottom-right indicate directions consistent with Equation~\eqref{velocity_components}. 
(b) Brightness profiles of the jets as a function of emission age $\tau$. 
Cyan and magenta markers represent the eastern and western jets, respectively. 
Broken and solid lines represent flux before and after applying the scaling factors in panel (c). 
Error bars are shown only for the corrected flux and indicate statistical uncertainties. 
(c) Scaling factors accounting for both projection and Doppler beaming effects. 
These are applied to the profiles shown as broken lines in panel (b).
}
 \label{jet_model_with_deconvolve}
\end{figure*}

Figure~\ref{jet_model_with_deconvolve}(a) shows the modeled jet emission locations, overlaid on Figure~\ref{obs_subpixel_deconv}(c). The approaching and receding jets are represented in blue and red, respectively. The image is primarily consistent with the model, indicating that the observed emission most likely originates from the jet components themselves. This supports the interpretation that the emission arises not from a compact region near $\sim10^{17}$~cm from the core, but from an extended structure spanning $\sim6 \times 10^{16}$~cm \citep{Khabibullin_2017}. The two knotty structures located symmetrically east and west of the core share the same elapsed time of $t_{\mathrm{age}} \sim 200$~days, suggesting that they originate from the same ejection event. Although most features align well with the model, a slight deviation is observed in the northwest region at $t_{\mathrm{age}} \sim 100$~days, which may result from the readout streak effect, as the CCD charge transfer direction aligns with this offset (see Table~\ref{used_data}).

Figure~\ref{jet_model_with_deconvolve}(b) shows the X-ray intensity along the jet as a function of emission age $\tau$, derived from the jet structure in Figure~\ref{jet_model_with_deconvolve}(a). 
The cyan and magenta curves represent the eastern and western jets, respectively. 
To extract the intrinsic brightness distribution, we apply correction factors that account for Doppler beaming, projection effects due to precession, and the spatial resolution of the image. 
These corrections follow the method of \citet{Bell_2011} (see Appendix~\ref{Projection and Beaming Correction Factors}), with the spatial resolution estimated from the RL-deconvolved image (see Appendix~\ref{Evaluating the Spatial Resolution of RL Deconvolution}).
The broken lines show the uncorrected flux, while the solid lines represent the flux after applying these corrections. 
The correction factors themselves are visualized as a scaling map in Figure~\ref{jet_model_with_deconvolve}(c), and were used to compute the solid-line curves in Figure~\ref{jet_model_with_deconvolve}(b).

The corrected brightness (solid lines) in Figure~\ref{jet_model_with_deconvolve}(b) indicates that for $\tau \lesssim 100$~days, the brightness includes contributions from the core. Between $\tau \sim 100$ and $\sim 200$~days, both jets show brightness variations of about an order of magnitude. The eastern jet remains nearly constant in brightness over $200 \lesssim \tau \lesssim 250$~days, whereas the western jet decreases by an order of magnitude. Beyond $\tau \sim 250$~days, the eastern jet gradually declines, while the western jet follows a more complex trend due to the overlap of multiple precessional components. The characteristics of this flux decay are discussed in Section~\ref{Jet Brightness Decay and Emission Mechanisms}.

\subsection{X-ray Spectral Properties of Jet Components}

To investigate the physical properties of the X-ray emitting jets, zeroth-order and first-order ($\pm$1) HEG spectra were analyzed.
Since the zeroth-order spectra have lower energy resolution than the HEG data, the spectral parameters were initially determined using the high-resolution HEG data.

The adopted model, \texttt{tbabs*(bvapec\_b + bvapec\_r + gauss)}, consists of thermal plasma emission from both jets and a fluorescent Fe\,\textsc{i} K$\alpha$ line at 6.4~keV (e.g., \cite{Marshall_2002}).  
\texttt{bvapec\_b} and \texttt{bvapec\_r} correspond to the eastern and western jets, respectively.  
Because the jets are nearly perpendicular to the line of sight, similar relativistic effects and plasma conditions were assumed.
The temperature, velocity broadening, and normalization of \texttt{bvapec\_b} were linked to those of \texttt{bvapec\_r}.   
The Fe\,\textsc{i} K$\alpha$ line has FWHM $< 1000~\mathrm{km~s^{-1}}$ \citep{Marshall_2002}, and the Gaussian width was fixed to $\sigma_E = 10~\mathrm{eV}$. 
Elemental abundances were fixed to solar values based on the \texttt{aspl} table \citep{Asplund_2009}, and the hydrogen column density was fixed at $1 \times 10^{22}~\mathrm{cm^{-2}}$ \citep{Marshall_2002}.

\begin{figure*}[ht!]
 \includegraphics[width=1\linewidth]{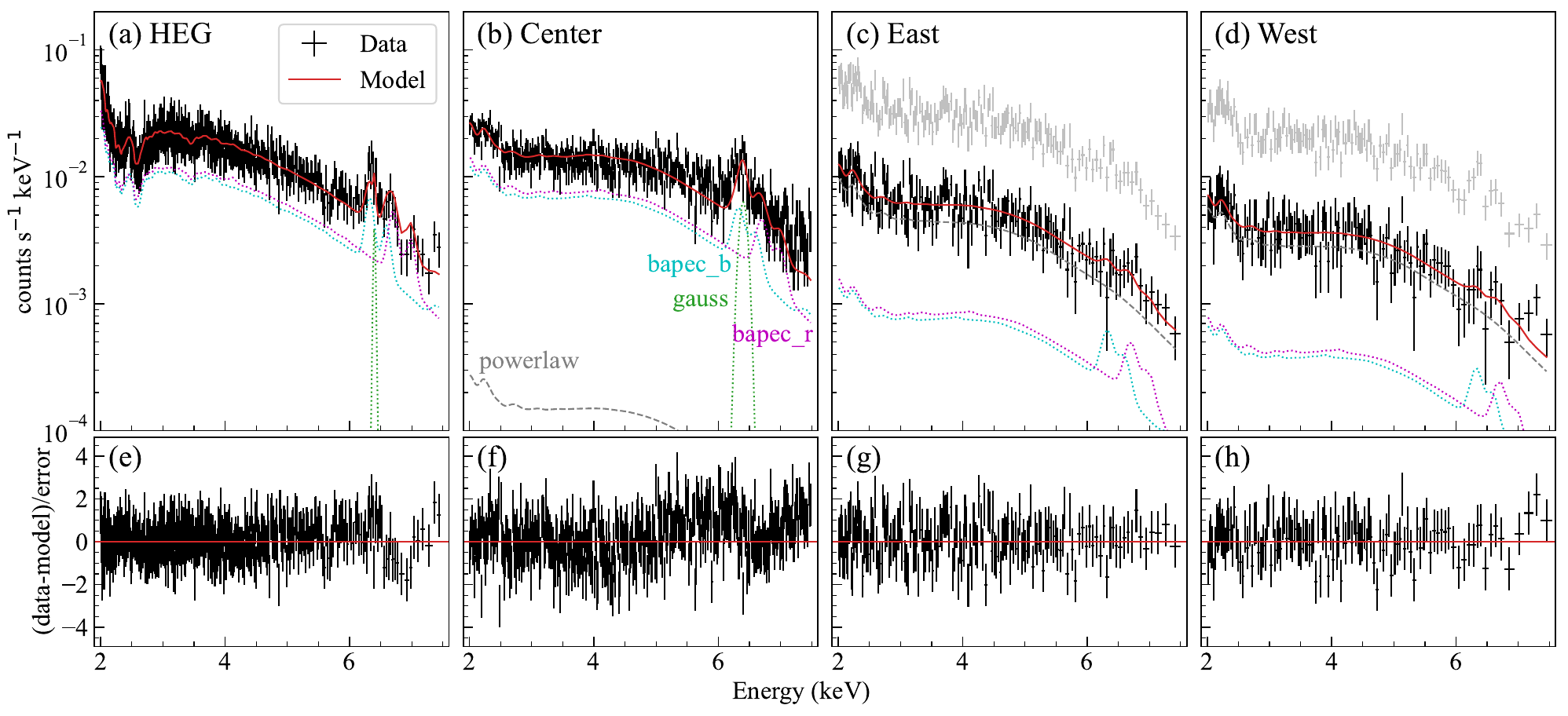}
 \caption{Spectral fits to HETG/HEG and zeroth-order data (2--7.5~keV) using an absorbed thermal plasma plus power-law model with a Gaussian Fe\,\textsc{i} K$\alpha$ component: \texttt{tbabs*(bvapec\_b + bvapec\_r + gauss + powerlaw)}.  
Zeroth-order fit was constrained using HEG spectra, with linked and fixed parameters listed in Table~\ref{fit_table_thermal_powerlaw}.  
(a--d) HEG ($\pm$1) spectra and zeroth-order spectra from Center, East, and West regions, respectively, as defined in Figure~\ref{obs_subpixel_deconv}.  
For East and West regions, both spectra before PSF contamination subtraction from Center region (shown in gray;  scaled by a factor of 4 for clarity) and after subtraction (shown in black) are plotted.  
Individual components --- \texttt{powerlaw}, \texttt{bvapec\_b}, \texttt{bvapec\_r}, and \texttt{gauss} --- are shown in gray, cyan, magenta, and green, respectively.  
(e--h) Residuals between data and model for corresponding regions shown in panels (a--d).
}
 \label{fit_thermal_powerlaw}
\end{figure*}

The fitting results for the HEG spectra are shown in Figure~\ref{fit_thermal_powerlaw}(a), and the corresponding parameters are summarized in Table~\ref{fit_table_thermal_powerlaw}.  
The best-fit redshifts, $z_\mathrm{b}$ and $z_\mathrm{r}$, are consistent with the Doppler shifts expected from the precession phase (see Figure~\ref{orb_vs_jet_phase}(b)).
The plasma temperature ($kT$) and Gaussian normalization ($A_\mathrm{gau}$) are consistent with previous studies (e.g., \cite{Marshall_2002, Medvedev_2019}).
These consistencies support the reliability of the HEG-based spectral model, which is subsequently applied to the zeroth-order data.

Because the East and West spectral regions are contaminated by PSF leakage from the Center region, the amount of leakage was quantified.  
The PSF map was integrated over each region (Center, East, and West), and the fractional contributions from the Center to the East and West were calculated to be $p_{\mathrm{center \to east}} \sim 0.130$ and $p_{\mathrm{center \to west}} \sim 0.114$, respectively.  
The estimated contamination was subtracted from the East and West spectra using the scaled spectrum from the Center region.

Figures~\ref{fit_thermal_powerlaw}(b--d) show the zeroth-order spectra from the Center, East, and West regions.
For the East and West regions, the spectra are plotted both before (gray) and after (black) subtraction of the Center spectrum.
Although the East and West spectra exhibit a weak Fe-K line feature prior to subtraction, it is primarily due to Fe-K photons (thermal and fluorescent) scattered from the Center region via the PSF.

To phenomenologically model the observed spectral features in the outer regions, the zeroth-order spectra were analyzed using a spectral model, \texttt{TBabs*(bvapec\_b + bvapec\_r + gauss + powerlaw)}.
Since the photon index in the Center region was not well constrained, it was fixed at $\Gamma = 1.7$ to ensure stability.
Normalization parameters of the plasma and Gaussian components ($A_\mathrm{bapec}$ and $A_\mathrm{gau}$) were left free, while all other parameters were fixed to the best-fit values from the HEG spectra.
The fitting results for the zeroth-order spectra are shown in Figures~\ref{fit_thermal_powerlaw}(b--d) and summarized in Table~\ref{fit_table_thermal_powerlaw}.
The spectral characteristics of the East and West regions and their possible interpretations are discussed in Section~\ref{Implications of Spectral Analysis}.

\begin{table*}[ht!]
\caption{Fit results for Figure \ref{fit_thermal_powerlaw}}
\begin{tabular*}{\textwidth}{@{\extracolsep{\fill}}lc|ccccc}
\toprule
Component & Parameter & HEG & Center & East & West \\
\midrule[\heavyrulewidth]
\texttt{TBabs} & $N_\mathrm{H}$ ($10^{22}\, \mathrm{cm^{-2}}$) & $1.0$ (fixed) & $^{\ast}$ & $^{\ast}$ & $^{\ast}$ \\\midrule
\multirow{4}{*}{\texttt{bapec\_b}}
&$kT$ (keV) & $8.7^{+1.9}_{-1.2}$ & $^{\ast}$ & $^{\ast}$ & $^{\ast}$ \\
& $z_\mathrm{b}$ & $0.056\pm 0.006$ & $^{\ast}$ & $^{\ast}$ & $^{\ast}$ \\
& ${\sigma_{v_\mathrm{jet}}}^a$ ($\mathrm{10^3\,km\, s^{-1}}$) & $2.7^{+1.4}_{-1.0}$ & $^{\ast}$ & $^{\ast}$ & $^{\ast}$ \\
&$A_\mathrm{bapec}$$^b$ ($\times 10^{-4}$) & $94.4^{+3.7}_{-3.4}$ & $20.2^{+0.6}_{-4.2}$ & $2.2^{+2.6}_{-2.2}$ & $1.1^{+2.2}_{-1.1}$ \\\midrule
\multirow{4}{*}{\texttt{bapec\_r}}
&$kT$ (keV) & $^{\dagger}$ & $^{\dagger}$ & $^{\dagger}$ & $^{\dagger}$ \\
&$z_\mathrm{r}$ & $-0.003^{+0.006}_{-0.005}$ & $^{\ast}$ & $^{\ast}$ & $^{\ast}$ \\
& ${\sigma_{v_\mathrm{jet}}}^a$ ($\mathrm{10^3\,km\, s^{-1}}$) & $^{\dagger}$ & $^{\dagger}$ & $^{\dagger}$ & $^{\dagger}$ \\
&$A_\mathrm{bapec}$$^b$ ($\times 10^{-4}$) & $^{\dagger}$ & $^{\dagger}$ & $^{\dagger}$ & $^{\dagger}$ \\\hline
\multirow{3}{*}{\texttt{gauss}}
& $E_\mathrm{line}$ (keV) & $6.4$ (fixed) & $^{\ast}$ & $^{\ast}$ & $^{\ast}$ \\
& $\sigma_E$ (eV) & $10$ (fixed) & $^{\ast}$ & $^{\ast}$ & $^{\ast}$ \\
& ${A_\mathrm{gau}}^c$ ($\times 10^{-6}$) & $9.8^{+8.8}_{-8.9}$ & $11.1^{+2.7}_{-2.5}$ & $0.0^{+1.5}_{-0.0}$ & $0.0^{+1.5}_{-0.0}$ \\\midrule
\multirow{2}{*}{\texttt{powerlaw}}
& $\Gamma$ & n/a & $1.7$ (fixed) & $2.0^{+0.3}_{-0.2}$ & $1.9^{+0.4}_{-0.2}$ \\
& ${A_\mathrm{pl}}^d$ ($\times 10^{-5}$) & n/a &$1.2^{+24.9}_{-1.2}$ & $52.8^{+16.5}_{-15.4}$ & $29.2^{+12.5}_{-13.2}$ \\\midrule
\multicolumn{2}{c|}{Energy Flux$^{e}$} & & & &  &  \\
\texttt{bapec\_b+bapec\_r} & $F_\mathrm{th}$& $186.7\pm 4.5$ & $40.1^{+1.1}_{-8.3}$ & $4.8\pm 4.8$ & $2.2^{+4.5}_{-2.2}$ \\
\texttt{powerlaw} & $F_\mathrm{pl}$ &  n/a & $0.0^{+8.4}_{-0.0}$ & $11.1^{+5.1}_{-4.7}$ & $7.4^{+2.7}_{-4.2}$ \\
\texttt{bapec\_b+bapec\_r+gauss+powerlaw} & $F_\mathrm{int}$ & $187.7^{+4.5}_{-4.6}$ & $41.6^{+0.7}_{-0.8}$ & $16.0\pm 0.6$ & $9.7\pm 0.5$\\\midrule
\multicolumn{2}{c|}{$C$-stat/d.o.f.} & $186/368$ & $453/368$ & $141/165$ & $98/123$ \\
\bottomrule
\end{tabular*}

\vspace{0.3cm}
\small
\textbf{Notes.}---
All errors are quoted at the 90\% confidence level throughout this table.
$^{a}$ Parameter range set to 200--5000~$\mathrm{km\,s^{-1}}$ to ensure numerical stability.
$^{b}$ \texttt{bapec} normalization: emission measure, defined as 
$\frac{10^{-14}}{4\pi [D_A(1+z)]^2} \int n_e n_\mathrm{H}\, dV$ 
($\mathrm{cm^{-5}}$), where $D_A$ is angular diameter distance, 
$z$ is redshift, and $n_e$, $n_\mathrm{H}$ are densities in 
$\mathrm{cm^{-3}}$.
$^{c}$ Gaussian normalization unit: $\mathrm{photons\, cm^{-2}\, s^{-1}}$.
$^{d}$ Power-law normalization unit: $\mathrm{photons\, keV^{-1}\, cm^{-2}\, s^{-1}}$ at 1~keV.  
$^{e}$ Energy flux unit (2--7.5~keV): $10^{-13}~\mathrm{erg\,cm^{-2}\,s^{-1}}$
Power-law component was not included in the HEG fit.
$^{\ast}$ Fixed to values from the HEG spectra.  
$^{\dagger}$ Linked to the corresponding parameter in \texttt{bapec\_b}.
\label{fit_table_thermal_powerlaw}
\end{table*}

\section{Discussion}\label{Discussion}

\subsection{Phase-Dependent Jet Visibility} \label{Phase-Dependent Jet Visibility}

The available HETG images allow investigation of how jet visibility varies with both orbital and precessional phases.
In Figure~\ref{all_hetg_flux_ratio}(a), Obs.~IDs~1940, 1941 \citep{migliari2005rapid}, and 15781 exhibit notably higher flux ratios compared to other epochs. These observations share a common feature: low HEG flux ($F_\mathrm{HEG}$; see Table~\ref{used_data}), which indicates that the core emission was likely suppressed. Supporting this interpretation, the subsequent observation Obs.~ID~1942, taken only a few days after Obs.~ID~1940, shows a lower ratio, likely due to stronger core emission during an off-eclipse orbital phase.  
Such high flux ratios are expected when two conditions are simultaneously satisfied: minimized Doppler beaming, which occurs when the jet lies close to the plane of the sky, and eclipse-induced suppression of the core emission (see also hard X-ray studies in the 18--60 keV band on phase-dependent flux variations; \cite{Cherepashchuk_2020}).

\subsection{Jet Brightness Decay and Emission Mechanisms}\label{Jet Brightness Decay and Emission Mechanisms}

To interpret the jet flux decay, the van der Laan model \citep{van_1966} provides a general framework for power-law decay driven by synchrotron radiation and adiabatic expansion, broadly applicable to astrophysical jets. \citet{Hjellming_1988} developed a detailed adiabatic expansion model specifically for SS~433, describing two distinct phases: an initial slowed expansion transitioning to free expansion under optically thin conditions, where lateral expansion dominates the luminosity decay. \citet{Bell_2011} reported that the radio flux decay in SS~433 is better described by an exponential profile. To investigate the X-ray radiative properties, which may involve multiple emission mechanisms, we analyze the X-ray intensity as a function of emission age.

The corrected brightness decay shown in Figure~\ref{jet_model_with_deconvolve}(b) is compared with previous studies. According to \citet{roberts2010structure}, who conducted a VLA observation in 2003 ($\phi_{\mathrm{prec}} \sim 0.47$), the corrected brightness of the eastern jet remained nearly constant over $200 \lesssim \tau \lesssim 250$~days, consistent with our X-ray findings. The subsequent exponential decay observed in the eastern X-ray jet also agrees with trends seen in the radio band (e.g., \cite{roberts2010structure,Bell_2011}). To quantify this decay, the corrected brightness was fit with an exponential decay function of the form $e^{-\tau / \tau'}$, yielding a decay timescale of $\tau' = 35.3 \pm 3.3$~days. This result suggests that the X-ray emission decays slightly more rapidly than radio emission ($\tau' = 55.9 \pm 1.7$~days; \cite{Bell_2011}), potentially reflecting differences in cooling mechanisms or interactions with the surrounding interstellar medium, such as synchrotron cooling in the X-ray regime.

\subsection{Correlation of X-ray and Radio Images}\label{Correlation of X-ray and Radio Images}

Understanding the relationship between X-ray and radio emission is key to studying jet dynamics and particle acceleration in SS~433. In jet emissions, X-rays may arise from thermal and/or non-thermal processes, while radio emission primarily results from synchrotron radiation. By comparing their spatial structures, we assess whether the observed flux densities and spectra are consistent with a power-law behavior.

\begin{figure}[ht!]
 \includegraphics[width=1\linewidth]{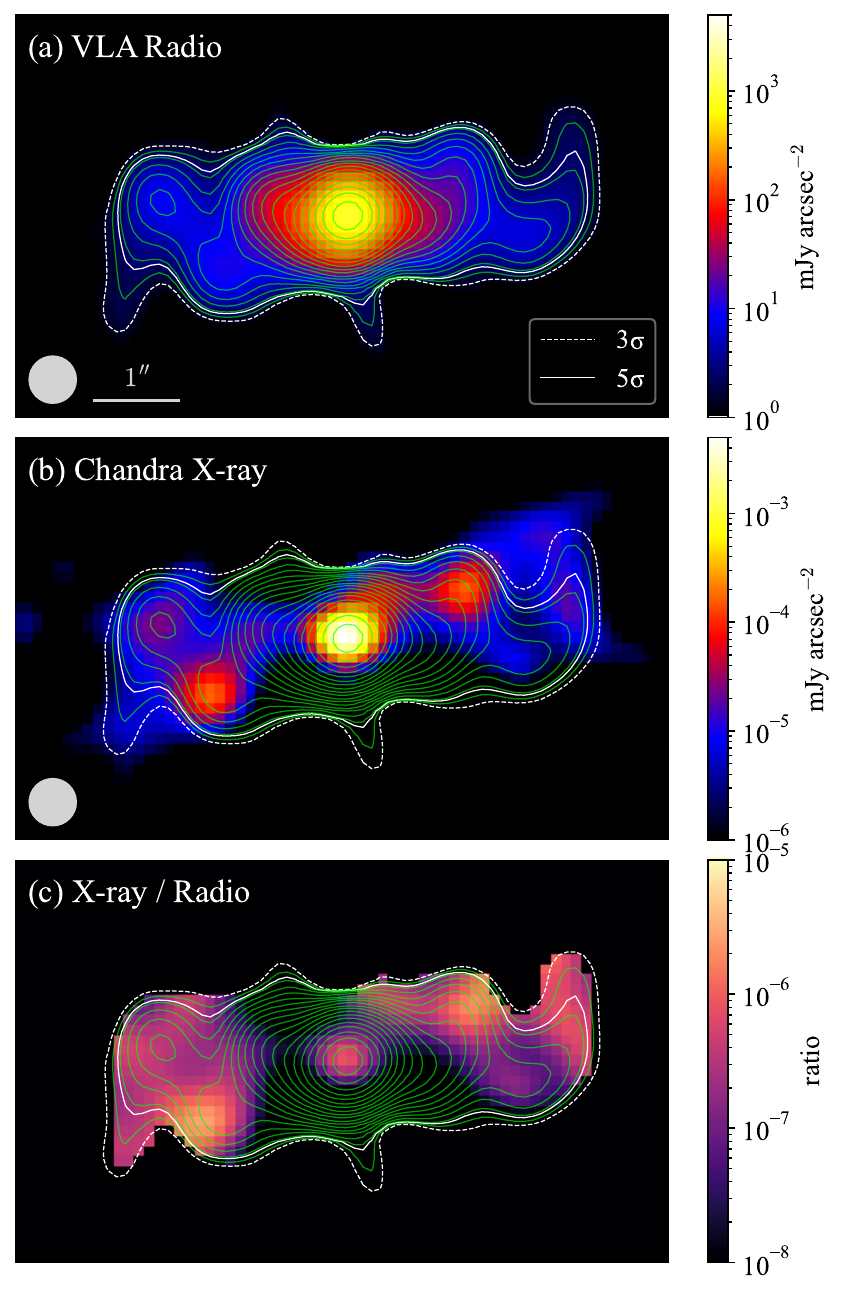}
 \caption{Comparison of SS 433's radio and X-ray jet structures.  
(a) VLA radio image with the beam size (FWHM = $0\farcs58$) shown in the bottom-left corner.  
Green contours start at $4\sigma$ of the RMS noise  
($\sigma = 0.18~\si{mJy.beam^{-1}}$, corresponding to $\sim 0.47~\si{mJy.arcsec^{-2}}$) and increase by factors of $\sqrt{2}$.  
White dashed and solid contours represent $3\sigma$ and $5\sigma$ significance levels in panel (a).  
(b) Chandra X-ray image (Figure~\ref{obs_subpixel_deconv}(c)), overlaid with contours from panel (a).  
The circle in panel (b) indicates the FWHM ($0\farcs58$), as calculated in Appendix~\ref{Evaluating the Spatial Resolution of RL Deconvolution}.  
The unit is converted as follows: \si{mJy.arcsec^{-2}} $= 1.35$~\si{photons.cm^{-2}.\textit{detector}.\textit{pixel}.\textit{size}^{-2}.s^{-1}}.  
The precession phases for panels (a) and (b) are $\phi_{\mathrm{prec}} \sim 0.47$ and $\sim 0.43$, respectively.  
(c) Panel (b) divided by panel (a), showing regions above the $3\sigma$ RMS threshold.
}
 \label{vla_vs_chandra}
\end{figure}

Figure~\ref{vla_vs_chandra}(a) shows the total intensity map derived from VLA observations (see Section~\ref{Very Large Array} for details). To visualize the signal relative to the background noise, contours are overplotted based on the RMS noise level. 
Although standard radio analyses typically focus on features above $5\sigma$, we include contours down to $3\sigma$ to highlight fainter but still reliable structures.
Figure~\ref{vla_vs_chandra}(b) shows the corresponding Chandra image (from Figure~\ref{obs_subpixel_deconv}(c)), overlaid with the same radio contours.
Both images exhibit comparable east--west extensions, with the X-ray emission aligning with significant radio features. For consistency in flux density comparisons, both images are expressed in units of \si{mJy.arcsec^{-2}}.

Between $\tau \sim 150$ and $\sim 200$~days, the X-ray knotty structures brighten by nearly an order of magnitude relative to the surrounding jet components (Figure~\ref{vla_vs_chandra}(a)), whereas in the radio band, the corresponding increase is only by a few times (Figure~\ref{vla_vs_chandra}(b)). This difference cannot be fully accounted for by Doppler beaming or overlapping precession phases, as these mechanisms would produce comparable enhancements in both bands if the emissions originated from the same electron population. \citet{migliari2005rapid} detected rapid X-ray variability on a timescale of a few~days, suggesting that transient processes such as internal shocks may influence the observed X-ray emission. These interactions likely drive localized energy release, contributing to the formation of the X-ray knotty structures. 

To explore the emission properties of the jet, the X-ray-to-radio flux ratio is examined.  
Figure~\ref{vla_vs_chandra}(c) shows the resulting ratio map, constructed by dividing the X-ray flux in panel (b) by the radio flux in panel (a). The brightness ratio of X-ray-to-radio emission in this region ranges from $10^{-7}$ to $10^{-5}$ at $\tau > 200$. Given a frequency range spanning $10^8$, the derived spectral index ($\alpha \sim 0.6$--$0.9$) is consistent with previous radio observations (e.g., \cite{Stirling_2004,Bell_2011}). This result also agrees with the spectral indices and flux ratios reported by \citet{Miller_2008}, and supports the interpretation that synchrotron radiation contributes to the jet emission observed in the X-ray band.
At $\tau < 200$, the flux ratio exhibits larger variations, ranging from $10^{-8}$ to $10^{-5}$, potentially indicating more complex emission conditions in this region.

\subsection{Implications of Spectral Analysis}\label{Implications of Spectral Analysis}

The spectral results presented in Figure~\ref{fit_thermal_powerlaw} and Table~\ref{fit_table_thermal_powerlaw} are used to investigate the physical origin of the observed spectral features.
The difference in intrinsic energy flux, $F_{\mathrm{int}}$, between the eastern and western jets was examined in the context of Doppler beaming. Observed photon energy is Doppler-shifted according to $E_{\mathrm{obs}} = D E_0$, where $D = [\gamma (1 - \beta_\ell)]^{-1}$ is the Doppler factor, and $E_0$ is the rest-frame energy. Photon flux density is affected by Doppler beaming and scales as $D^{n + \alpha}$, where $\alpha = \Gamma - 1$.  The exponent $n$ depends on the jet morphology: $n = 2$ for continuous jet flows and $n = 3$ for discrete or isolated jet components \citep{Sikora_1997}. For SS~433, the jets are typically modeled with $n = 2$, based on VLA radio observations \citep{roberts2010structure}.

Since the East and West regions cover more than one full precession cycle of the jet motion (see Figure~\ref{jet_model_with_deconvolve}(a)), it is reasonable to compute the average Doppler beaming factor over a full cycle. This calculation assumes uniform emission and symmetry between the east and west jets.
The intrinsic energy flux averaged over an energy bin $\Delta E$ can be expressed as
\begin{equation}
\langle F_\mathrm{int} \rangle \propto \int_0^{2\pi} \int_E^{E+\Delta E} D_\psi^{n + \alpha} \, D \, dE \, d\psi 
= \Delta E \int_0^{2\pi} D_\psi^{n + \alpha + 1} \, d\psi.
\end{equation}
Under this simplified model, the expected flux ratio between the eastern and western jets is $\langle F_{\mathrm{int,east}} \rangle / \langle F_{\mathrm{int,west}} \rangle \sim 1.5$ for $n=2$, and $\sim 1.6$ for $n=3$.
The observed flux ratio between the eastern and western regions is $F_{\mathrm{int,east}} / F_{\mathrm{int,west}} = 1.6 \pm 0.1$ (90\% confidence level), which is consistent with the values expected from relativistic Doppler boosting.

The spectral characteristics of the central and outer regions can be interpreted as follows.  
The central spectrum is well described by a thermal component ($F_\mathrm{th}$) and a neutral Fe\,\textsc{i} K$\alpha$ line ($A_\mathrm{gau}$).  
These features suggest that the emission in the central region is predominantly thermal.  
In contrast, while $F_\mathrm{th}$ and $A_\mathrm{gau}$ are consistent with zero at the 90\% confidence lower limits in the outer regions, the power-law component ($A_\mathrm{pl} > 0$) is constrained to be positive at the same confidence level.
This result indicates that the presence of the power-law component cannot be ruled out in the outer regions.
The physical origin of this component remains an open question, but its spatial coincidence with the radio jets (see Section~\ref{Correlation of X-ray and Radio Images}) suggests a possible association with non-thermal electrons.

To gain deeper insight into the physical nature of the outer jet emission, several avenues for future investigation can be considered.
As discussed in Section~\ref{Phase-Dependent Jet Visibility}, deep observations during orbital and precessional phases where the jets are more favorably oriented for detection hold promise for future studies.
In addition, as suggested by \citet{migliari2002iron} and \citet{Khabibullin_2017}, the observed spectra of the outer regions include multiple Doppler-shifted components originating from different precessional phases, blended due to the limited spatial resolution of current instruments (see also Figure~\ref{jet_model_with_deconvolve}(a), where multiple jet components are likely visible in the outer regions).
Incorporating such effects into spectral modeling will be important for accurately characterizing the emission.
Nutation motion \citep{katz_1982} may also contribute to the observed spectral variability and warrants further investigation.
High-resolution spectroscopy from ongoing missions such as the X-Ray Imaging and Spectroscopy Mission (XRISM; \cite{Tashiro_2022}) is expected to clarify physical mechanisms---including precession and nutation---and their influence on spectral features, offering insights into the formation and evolution of relativistic jets in SS~433.

\section{Conclusion} \label{Conclusion}

We analyzed the arcsecond-scale spatial structures of the relativistic jets in SS~433 using X-ray data from the Chandra HETG observations.    
The dataset includes 24 observations spanning from 1999 to 2024, totaling $\sim$~850~ks in exposure and covering a broad range of orbital and precessional phases.
The observed images consistently show enhanced X-ray emission along the east--west direction, indicating the persistent presence of arcsecond-scale jet structures.  
These features become more discernible when the core emission is suppressed, particularly during orbital eclipse and when the jet axis is oriented close to the plane of the sky, reducing Doppler beaming.

To enhance spatial resolution, the RL deconvolution was applied. The 2014 data revealed, for the first time, two knotty structures with a helical morphology.
The deconvolved image is generally consistent with the kinematic precession model inferred from radio observations.
The comparison with VLA radio observations from 2003, taken at a similar precessional phase, revealed comparable spatial scales. Although the spectral indices are generally consistent between the X-ray and radio bands, notable differences emerge at the knotty structures.

Spatially resolved spectral analysis of the zeroth-order data indicates that any Fe emission lines in the outer regions are below the detection threshold after accounting for the core contribution using a PSF model. 
Combined with the morphological similarity to the radio jets, this suggests that the arcsecond-scale X-ray emission likely arises from a non-thermal component.
These findings highlight the importance of future X-ray missions with high energy and spatial resolution (e.g., XRISM), combined with simultaneous radio observations, to further clarify the jet’s emission mechanisms.

\begin{ack}
We are grateful to the referee for their constructive feedback, which significantly improved the quality of this paper, particularly the spectral analysis.  
We also thank the editor for their careful handling of the review process.  
Insightful discussions with the SS~433 XRISM team (PI: Dr.~Megumi Shidatsu) and Dr.~Taro Kotani were invaluable throughout the development of this study.
This research made use of archival data from the Chandra X-ray Observatory and the Karl G. Jansky VLA, operated by the NRAO under cooperative agreement with the NSF.  
We performed X-ray image analysis using CIAO and subpixel PSF simulations with MARX, whose modeling capabilities were essential for the high-resolution imaging presented in this study.  
Spectral fitting was carried out using XSPEC, and radio data reduction was conducted with CASA.  
We acknowledge the developers and calibration teams of these tools for their indispensable contributions.  
This research was supported by JSPS KAKENHI Grant Numbers 24KJ2067, 22K20386, 22H01272, 23K22543, and 24K00672.
\end{ack}

\appendix

\section{Optimal Number of Iterations for the RL Deconvolution}\label{Optimal Number of Iterations for the RL Deconvolution}
The number of iterations in the RL deconvolution impacts both the quality of image restoration and the degree of noise amplification. Selecting the optimal number of iterations is challenging, as it depends on various factors including the region of interest, statistical fluctuations, and image characteristics. To address this, we adopt a semi-quantitative approach using the reduced chi-squared value as a convergence criterion, following \citet{Marchenko_2017}.

The reduced chi-squared for the $r$-th iteration is defined as
\begin{equation}
\chi^{2,(r)}_{\text{red}} = \frac{1}{d} \sum_j \frac{\left( N_j - (W^{(r)} \ast P)_j \right)^2}{\sigma_j^2},
\label{chi_eq}
\end{equation}
where $j$ is the pixel index, and $d$ is the number of degrees of freedom. The quantities $W$, $N$, and $P$ are the same as those defined in Equation~\eqref{rl_eq}.
The standard deviation $\sigma_j$ is defined as
\begin{equation}
\sigma_j = 
\begin{cases} 
\sqrt{N_j} & (N_j \geq 5) \\
1 + \sqrt{N_j + 0.75} & (N_j < 5)
\end{cases},
\end{equation}
to account for Poisson statistics in low-count data, following \citet{Gehrels_1986}.

\begin{figure}[ht!]
 \includegraphics[width=1\linewidth]{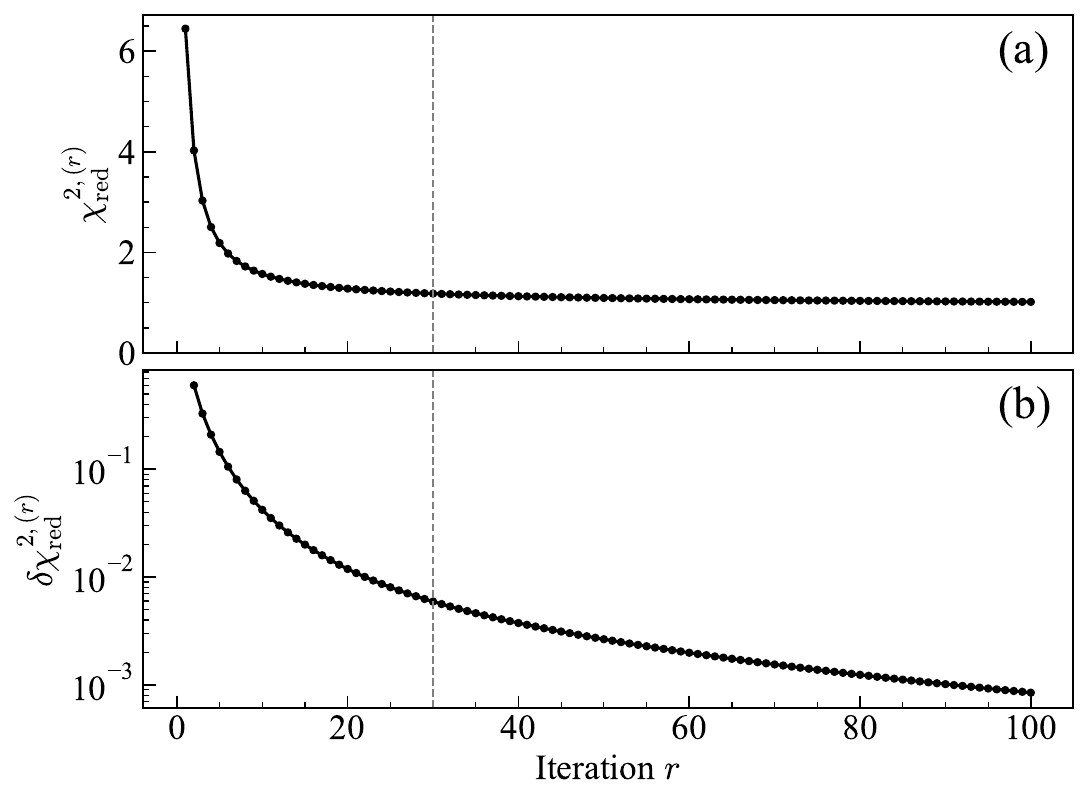}
 \caption{Reduced chi-squared $\chi^{2,(r)}_{\text{red}}$ as a function of RL iterations for Obs.~ID~15781.  
(b) Absolute rate of change $\delta\chi^{2,(r)}_{\text{red}}$ with respect to iterations.  
At iteration 30 (dashed line), $\chi^{2,(30)}_{\text{red}} = 1.18$, and $\delta\chi^{2,(30)}_{\text{red}} = 0.0057$, following the convergence criterion described by \citet{Marchenko_2017}.
}
 \label{rl_reduced_chi_ratio}
\end{figure}

To avoid bias from noise-dominated regions, the analysis was restricted to a rectangular region of 73$\times$45 pixels (in 1/4 pixel units), which focuses on the jet structures while excluding extraneous areas.
The reduced chi-squared was evaluated in increments of 10 iterations to track convergence trends. The absolute rate of change of $\chi^{2,(r)}_{\text{red}}$ is defined as
\begin{equation}
\delta \chi^{2,(r)}_{\text{red}} = \frac{|\chi^{2,(r)}_{\text{red}} - \chi^{2,(r-1)}_{\text{red}}|}{\chi^{2,(r)}_{\text{red}}}.
\end{equation}

Figure~\ref{rl_reduced_chi_ratio}(a) shows the evolution of reduced chi-squared value ($\chi^{2,(r)}_{\text{red}}$) as a function of iterations for Obs.~ID~15781, and Figure~\ref{rl_reduced_chi_ratio}(b) shows the rate of change $\delta \chi^{2,(r)}_{\text{red}}$. Iterations were continued until $\delta \chi^{2,(r)}_{\text{red}}$ fell below 10$^{-2}$, a threshold ensuring sufficient convergence while avoiding overfitting and noise amplification. Based on this criterion, the analysis identified 30 iterations as the optimal stopping point, balancing restoration quality with noise suppression and mitigating over-iteration artifacts.

\section{Light-Travel-Time Mapping of Jet Emission in SS 433}\label{Light-Travel-Time Mapping of Jet Emission in SS 433}

While the kinematic model in Equation~\eqref{velocity_components} provides a three-dimensional description of the jet velocity and geometry, it does not inherently account for the time delay associated with photon travel. This light-travel-time effect becomes important when projecting the model onto the two-dimensional sky plane, where photons from forward- and backward-moving components arrive at the observer at different times due to their line-of-sight separation. Equation~\eqref{tau_vs_tage_eq} captures this effect by relating the intrinsic birth epoch $t_{\mathrm{age}}$ of the jet material to the observed emission age $\tau$, based on the geometric treatment presented in \citet{roberts2010structure}. In this appendix, we visualize this relationship to clarify how the projection depends on the spatial orientation and depth of the jet.

\begin{figure}[ht!]
 \includegraphics[width=1\linewidth]{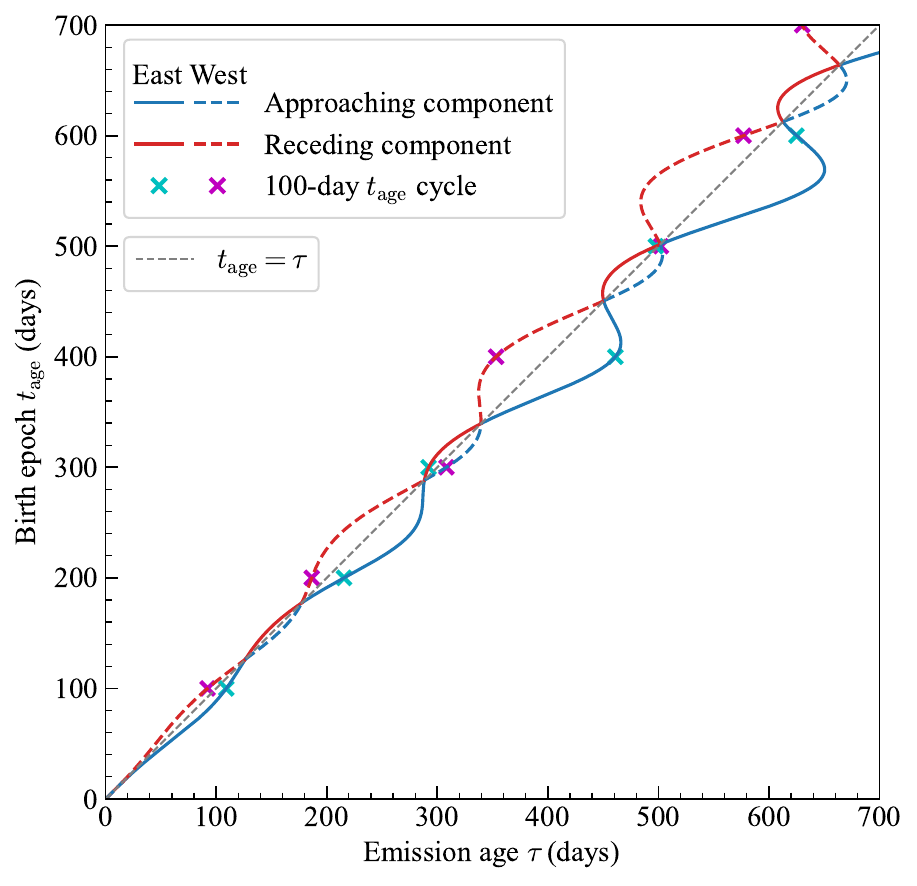}
 \caption{Light-travel-time effects in SS~433 calculated from Equation~\eqref{tau_vs_tage_eq}, following the formulation of \citet{roberts2010structure}.  
Start point $t_{\mathrm{age}} = \tau = 0$ is set to precession phase of Obs.~ID~15781.  
Vertical axis shows birth epoch $t_{\mathrm{age}}$ of jet material, and horizontal axis shows emission age $\tau$.  
Solid and dashed lines represent eastern and western jets, respectively.  
Blue and red indicate approaching and receding jet components.  
Cyan and magenta marks represent 100-day intervals of $t_{\mathrm{age}}$, plotted to illustrate correspondence with Figure~\ref{jet_model_with_deconvolve}.  
Gray dashed line indicates identity line $t_{\mathrm{age}} = \tau$.
}
 \label{tau_vs_tage}
\end{figure}

Figure~\ref{tau_vs_tage} shows the relationship between the birth epoch $t_{\mathrm{age}}$ and the emission age $\tau$ for the eastern and western jets in SS~433, illustrating the light-travel-time effect. Although $t_{\mathrm{age}}$ is uniquely defined for each component, multiple components may share the same $\tau$, meaning they appear simultaneously to the observer despite being ejected at different times.

The blue and red lines represent the approaching and receding components, respectively. By definition, $t_{\mathrm{age}} = \tau$ corresponds to emission directed perpendicular to the line of sight. For approaching components, regions with $t_{\mathrm{age}} < \tau$ indicate that the observer sees radiation from farther along the jet, corresponding to more elapsed time since ejection. This is a consequence of the light-travel-time effect, which causes the observer to see more evolved emission than the material's intrinsic age would suggest.

\section{Projection and Beaming Correction Factors}\label{Projection and Beaming Correction Factors}

To investigate the brightness distribution of the jets while mitigating geometric and relativistic biases---such as projection effects and Doppler beaming---we use a model-based correction scheme. 
By incorporating known jet geometry and kinematics, this approach allows us to estimate the underlying radiative properties more accurately.
Such corrections have been commonly applied in radio studies of SS~433 (e.g., \cite{Bell_2011, Marti_2018}).

In the following, we apply this methodology to our X-ray data and describe the computation of projection and beaming correction factors, following the framework proposed by \citet{Bell_2011}.
In this model, each jet component $q$ is assumed to be emitted at uniformly spaced time intervals in $t_{\text{age}}$, and its sky position is calculated by Equation~\eqref{ra_dec_position}. 
The correction factor $C_p$ at the position of component $p$ is defined as
\begin{equation}
    C_p = \frac{1}{C_0} \sum_{q} D_q^{n + \alpha} \exp\left[ -\frac{(\Delta R_{pq})^2}{2 \sigma^2} \right],
    \label{correction_eq}
\end{equation}
where the summation approximates an integral over the continuously distributed jet components. Here, $\Delta R_{pq}$ is the angular separation between components $p$ and $q$, and $\sigma = 0\farcs246$ represents the spatial resolution derived from the RL-deconvolved image (see Appendix~\ref{Evaluating the Spatial Resolution of RL Deconvolution}).

The Doppler factor $D_q$ is computed for each jet component $q$ (Note that $q$ is an index representing jet components and does not correspond directly to the precession phase $\psi$.). 
We adopt $n = 2$ and $\alpha = \Gamma - 1 = 0.9$, where $\Gamma = 1.9$ is the photon index obtained as the best-fit value from the spectral fitting (see Table~\ref{fit_table_thermal_powerlaw}). The normalization constant $C_0$ ensures that $C_p = 1$ at the jet core. The inverse correction factor, $1/C_p$, is referred to as the scaling factor, which is multiplied by the measured flux to correct for projection and beaming effects. The resulting scaling map is shown in Figure~\ref{jet_model_with_deconvolve}(c).

\section{Evaluating the Spatial Resolution of RL Deconvolution}\label{Evaluating the Spatial Resolution of RL Deconvolution}

The RL deconvolution is a method to infer the true sky brightness distribution that best explains the observed image given the instrumental PSF. However, the resulting image should be regarded as a plausible approximation that reproduces the observed PSF-convolved data, rather than a definitive reconstruction of the true sky. For quantitative analysis, especially when comparing with physical models, it is important to characterize the effective spatial resolution of the deconvolved image. In this appendix, we evaluate the resolution based on the $1 \sigma$ extent of the PSF before and after applying the RL method, following the approach of \citet{Sakai_2024}.

\begin{figure}[ht!]
 \includegraphics[width=1\linewidth]{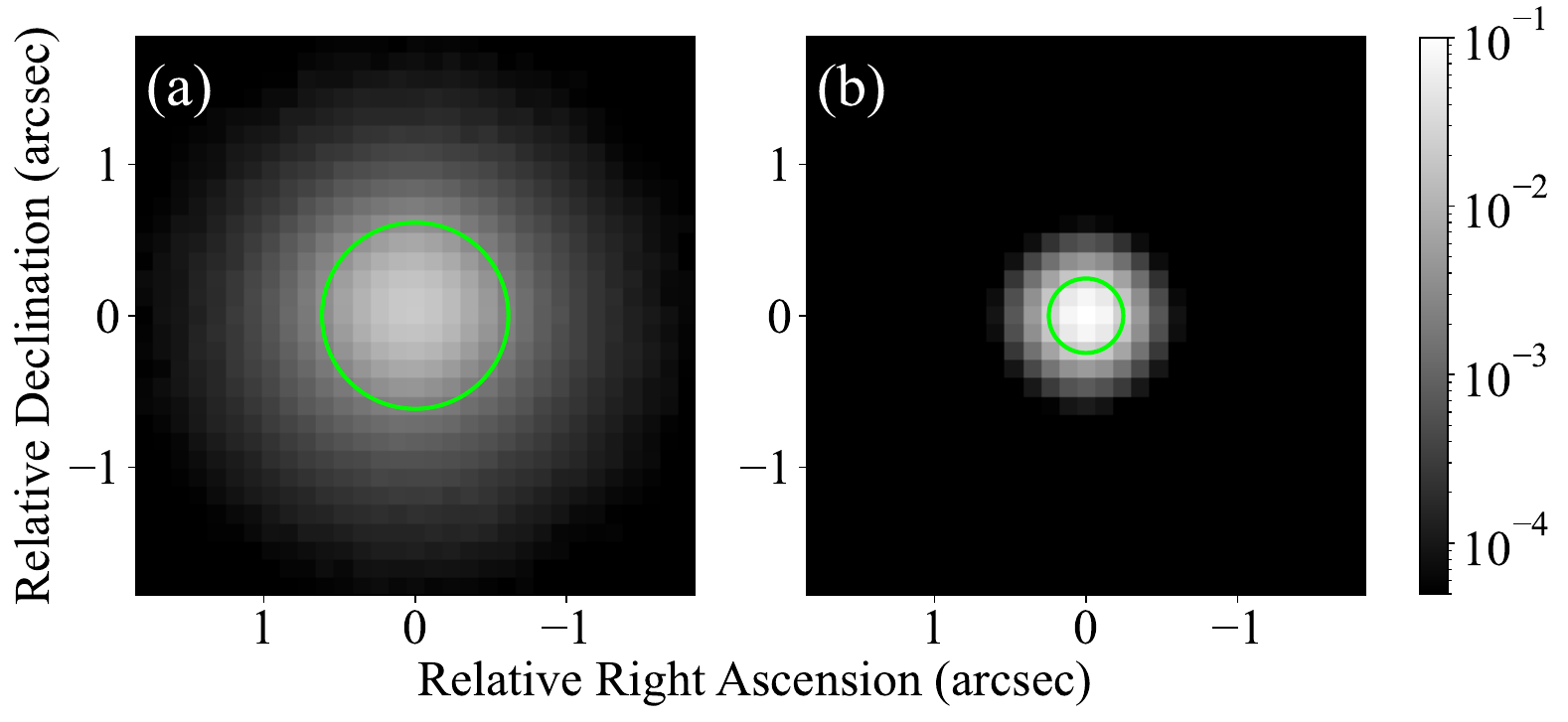}
 \caption{(a) Zeroth-order simulated PSF at 1/4 pixel scale and 3.7~keV for Obs.~ID~15781, normalized to an integral of 1.  
(b) PSF after 30 iterations of the RL method.  
Circles in both panels indicate $1\sigma$ confidence interval, based on resolution estimation approach described in \citet{Sakai_2024}.
}
 \label{psf_resolution}
\end{figure}

Figure~\ref{psf_resolution}(a) displays the zeroth-order PSF at a 1/4 pixel scale, simulated using \texttt{simulate\_psf} in CIAO for Obs.~ID~15781 with a monochromatic energy of 3.7~keV. After applying 30 iterations of the RL method, the deconvolved PSF is shown in Figure~\ref{psf_resolution}(b). In both panels, the grayscale indicates the probability density, with circles marking the $1 \sigma$ level.

The $1\sigma$ extent is $0\farcs615$ for the original PSF and $0\farcs246$ for the RL-deconvolved PSF. The $\sigma$ value of $0\farcs246$ represents the effective spatial resolution of the RL-deconvolved image and is used for calculating the correction factors described in Appendix~\ref{Projection and Beaming Correction Factors}.



\bibliographystyle{apj}
\bibliography{main}






\end{document}